\documentclass[conference]{IEEEtran}
\IEEEoverridecommandlockouts

\usepackage{cite}
\usepackage{amsmath,amssymb,amsfonts}

\usepackage{graphicx}
\usepackage{textcomp}
\usepackage{xcolor}
\usepackage{algorithm} 
\usepackage{algpseudocode} 
\usepackage[inline]{enumitem}
\usepackage{listings}
\usepackage[switch]{lineno}
\usepackage{todonotes}
\def\BibTeX{{\rm B\kern-.05em{\sc i\kern-.025em b}\kern-.08em
    T\kern-.1667em\lower.7ex\hbox{E}\kern-.125emX}}
\begin{document}

\title{Large-Scale Microtask Programming}

\author{
\IEEEauthorblockN{Emad Aghayi}
\IEEEauthorblockA{\textit{Department Of Computer Science, George Mason University, Fairfax, VA, eaghayi@gmu.edu}}
}

\IEEEoverridecommandlockouts
\IEEEpubid{\makebox[\columnwidth]{  \hfill} \hspace{\columnsep}\makebox[\columnwidth]{ }}

\maketitle
\IEEEpubidadjcol


\section{Introduction}

Crowdsourced software engineering offers many opportunities for reducing time-to-market, producing alternative solutions, employing experts, learning through work, and democratizing participation in software engineering. There are several types of crowdsourced software engineering. One of the oldest and most common is open source software development. Another approach is
competition-based crowdsourcing, where platforms such as TopCoder have increasingly become very popular with over 1,500,000 users.

A more recent form of crowdsourced software engineering is microtask programming.
Microtask programming decontextualizes work into self-contained microtasks, reducing the context necessary to onboard onto a software project and thereby decreasing joining barriers. At the same time, it may reduce the time to market for completing software work through parallelism.
Several prior systems have explored approaches for microtasking programming work, using either manual or automatic approaches for decomposing programming tasks into microtasks. Manual approaches rely on a developer~\cite{codeon2017} or client to author each microtask~\cite{lasecki2015apparition}. Microtask programming environments can reduce onboarding barriers through preconfigured web-based environments, such as Codepilot, CrowdCode, and Collabode\cite{CodePilot:2017,Collabode:2011,latoza2018microtask}. 

However, existing approaches for crowdsourced software development have significant limitations.
Open-source software development and competition-based approaches suffer from onboarding barriers, both technical and social. Although there are countless examples of successful open-source projects, onboarding challenges for newcomers can make it difficult to quickly onboard new developers and dissuade casual contributors. 
Microtask programming approaches can reduce onboarding barriers, both by offering a preconfigured environment as well as by enabling developers to do programming work with less prior knowledge or awareness of the complete project. But existing approaches are still limited in their support for design and architecture activities necessary to scale to larger software projects. Moreover, decontextualizing programming work is hard, bringing with it many challenges in doing it effectively. For example, conflicts may occur when two crowd workers make conflicting assumptions, necessitating approaches to reduce or repair conflicts as they occur~\cite{latoza2018microtask}.  

In my work, I have been exploring new ways to increase the \textit{scale} of microtask programming. I have developed a new behavior-driven development approach to microtask programming and conducted a series of studies to investigate the costs and benefits of microtask programming. In future work, I expect to continue to work to scale microtask programming to larger and more complex projects.



\begin{figure*}
\includegraphics[width=\textwidth,width=18cm, height= 8cm,keepaspectratio,clip]{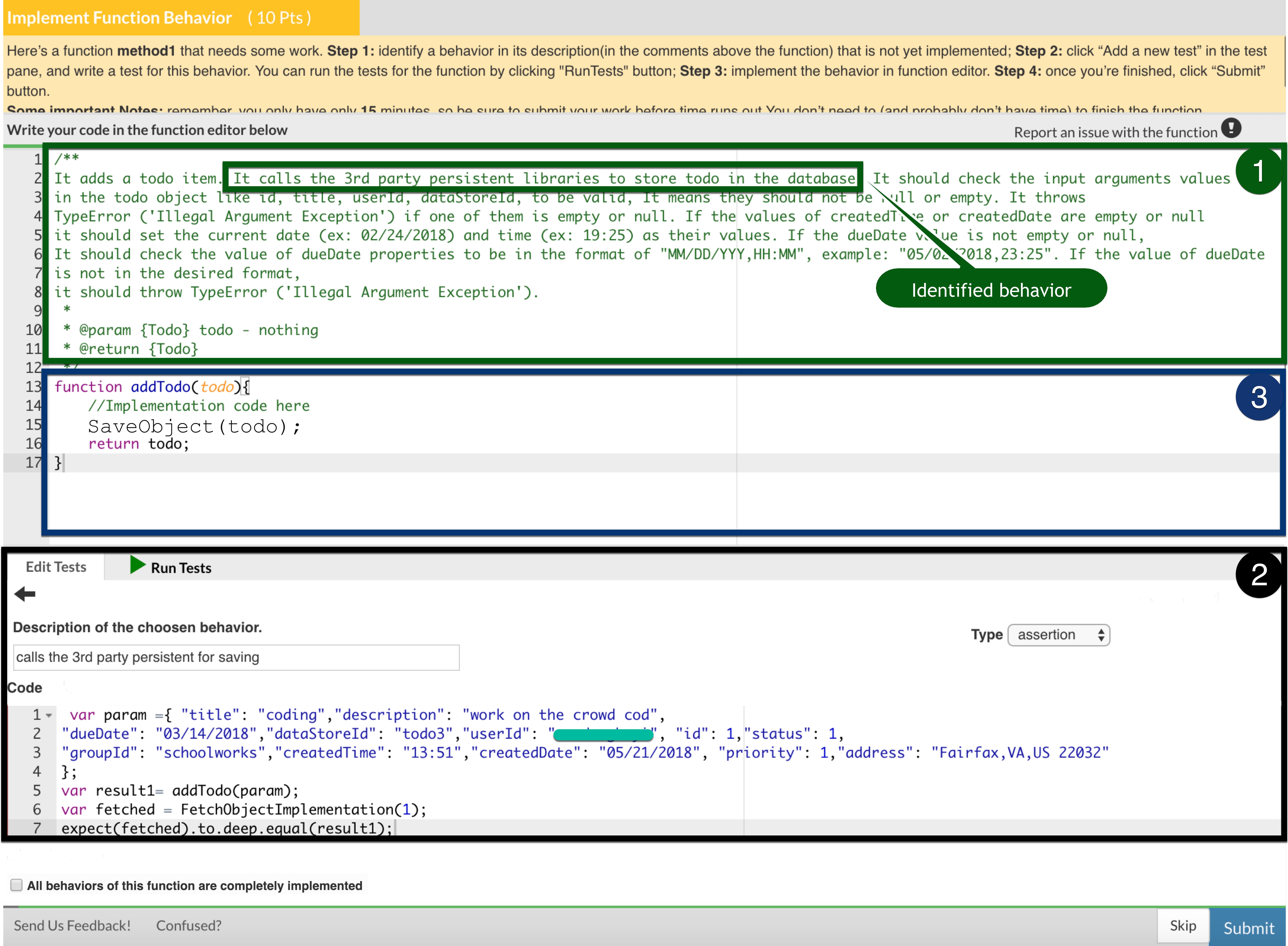}
\centering
\caption{In our behavior-driven microtask programming approach, developers complete microtasks where they (1) identify a behavior, (2) write a test for the behavior, and (3) implement the behavior~\cite{aghayi2019crowdsourced}. }
~\label{fig:workflow}
\end{figure*}

\section{Crowdsourced Behavior-Driven Development}

To make microtask programming more efficient and reduce the potential for conflicts between contributors, I developed a new behavior-driven approach to microtasking programming. In our approach, each microtask asks developers to identify a behavior behavior from a high-level description of a function, implement a unit test for it, implement the behavior, and debug it. It enables developers to work on functions in isolation through high-level function descriptions and stubs.

In addition, I developed the first approach for building microservices through microtasks. Building microservices through microtasks is a good match because our approach requires a client to first specify the functionality the crowd will create through an API. This API can then take the form of a microservice description. A traditional project may ask a crowd to implement a new microservice by simply describing the desired behavior in a API and recruiting a crowd. We implemented our approach in a web-based IDE, \textit{Crowd Microservices}\footnote{https://youtu.be/mIn2EOqsDYw} (Fig.~\ref{fig:workflow}). It includes an editor for clients to describe the system requirements through endpoint descriptions as well as a web-based programming environment where crowd workers can identify, test, implement, and debug behaviors (Figure \ref{fig:workflow}). The system automatically creates, manages, assigns microtasks. After the crowd finishes, the system automatically deploys the microservice to a hosting site.\par

\textit{Study 1: Feasibility.} To evaluate the feasibility of this approach, we conducted a small study where 9 developers together worked to build a microservice. The results were promising. Participants submitted their first microtask less than half an hour after beginning, successfully submitted 350 microtasks, implemented 13 functions and 36 tests, completed microtasks in a median time under 5 minutes, correctly implemented 27 of 34 behaviors, and together implemented most of a functioning ToDo microservice~\cite{aghayi2019crowdsourced}. 

\textit{Study 2: Comparing microtask programming to traditional development.} To directly compare traditional programming to microtasked programming, we conducted a controlled experiment. Twenty-eight developers worked either on traditional programming tasks, described through issues, or programming microtasks. We found that, compared to traditional software development, microtasking had important advantages in reducing onboarding time and time-to-market and, surprisingly, in increasing the quality of code and individual developer productivity.\par

\textit{Study 3: Using microtask programming in industry.} Our early studies were conducted entirely in artificial contexts, using artificial tasks and developers recruited specifically to work in the study. To examine the potential for using microtask programming in industry, we partnered with NTT, a large telecommunication company, to conduct a study of microtask programming within a real software project. We found that a microtask programming approach was successful in implementing and testing a project with 8000 lines of code in 14 functions. We also found that developers took time to understand the new concepts in microtask programming approach and be productive. We also found the value of having dedicated developers responsible for managing microtask projects.  \par


\section{Future Work}

My studies have offered initial evidence that microtask programming can be effective in small crowds with a few developers. But much of the promise of microtasking comes from large crowds, and there exists a direct relationship between the number of independent tasks and the parallelism in microtask programming which may reduce time to market. 
However, there are a number of significant challenges in scaling microtask programming to larger crowds. \par

%
To date, the largest crowd we have used is only 9 developers. To begin to examine microtask programming at scale, we are planning to conduct a virtual hackathon with around 100 developers. 
Based on our findings, we will develop new techniques to scale microtask programming and more fully encompass software development work. 

One direction we expect to pursue is with team organization. In our current approach, clients specify a microservice and crowd developers work in a flat organization to complete microtasks. This organization may not work well for all projects. For instance, some projects have security considerations, like private APIs that they do not want to expose to the public. Or projects may benefit from more experienced team leads who tackle more complex tasks or help less experienced developers when they get stuck. We will explore ways to make better use of more experienced developers in crowds through defining separate roles for crowd workers.
TopCoder is one successful example of this hybrid collaboration, and we will explore ways to adapt these ideas within microtask programming~\cite{Stol2014:TwoCompany}.\par


\section*{Acknowledgment}
This work was supported in part by the National Science Foundation under grants CCF-1414197 and CCF-1845508.

\bibliographystyle{IEEEtran}
\bibliography{FUU}

\end{document}